\documentclass[preprint,10pt,showpacs,prl,onecolumn]{revtex4}
\usepackage{amsmath}
\usepackage{amssymb}

\input{tcilatex}

\begin{document}

\preprint{HEP/123-qed}
\title[Exchange interaction in band theory ]{The exchange coupling and spin
waves in metallic magnets: removal of the long-wave approximation}
\author{V.P.Antropov}
\affiliation{Ames Lab, Ames, IA, 50011}

\begin{abstract}
A well-known connection between the magnetic susceptibility and the
effective exchange parameter is analyzed. It is shown that all current
computational schemes use a long-wave approximation which is suitable only
for localized moments systems. A corresponding smallness parameter is
derived in real and recipical space. General 'inverse susceptibility'
approach is combined with a multiple scattering theory and applied for the
studies of elementary metals Fe, Ni and Gd. It is shown that the proposed
approach significantly improves the description of the exchange coupling
between nearest atoms, the spin wave spectrum at large wave vectors for the
itinerant degrees of freedom in metallic magnets. A consistent usage of this
method leads to the proper description of magnetic short-range order effects.
\end{abstract}

\maketitle

The exchange coupling is one of the most important fundamental interactions
in magnets. Knowledge of the parameters of this interaction facilitates the
description of numerous properties of magnets and provides a critical test
of the applicability of different models. In addition, the current applied
research in the area of magneto-recording, spintronics and magnetotransport
phenomena requires the knowledge of the effective exchange coupling between
atoms or layers\cite{RESIST}. Therefore reliable calculations of the
parameters of this coupling are especially important both from fundamental
and applied points of view. Many computational expressions for magnets with
different magnetic orderings were proposed\cite%
{MORIYA,LIU,KORENMAN,OGU,LIXT,JMMM,REVIEW}. However, most of them were
derived only for localized spin systems at equilibrium. In this treatment
the atomic magnetization is directly proportional to the effective
'exchange' field (localized approach). In real non-equilibrium state these
directions are always different due to coexistence of localized and
itinerant degrees of freedom. Any model assumption about this non locality
restricts certain degrees of freedom and affects the spin-spin correlation
function and hence the dynamic and thermal properties. For instance, the
local approach implemented in Ref.\cite{KORENMAN} directly leads to the
concept of a strong short range magnetic order which by no means is a
universal behavior of magnets. It is desirable that both short-range (large $%
\mathbf{q}$) and long-range (small $\mathbf{q}$) fluctuations be present in
the general theory in a consistent fashion at any finite temperatures,
whereas the perturbation theory\cite{KORENMAN,OGU,LIXT,JMMM,REVIEW} is used
only at small temperatures (spin wave (SW) stiffness calculations). In
addition, for weak magnets or magnets with some weakly magnetized regions
the non-spherical parts of the effective magnetic field must be included.

In this paper starting from the general principles I provide a critical
analysis of the approximations used in the theory of exchange coupling, and
discuss an opportunity to calculate the parameters of such coupling
rigorously without specific assumptions about the range of magnetic order or
any approximations about the form of magnetization density.

In a nonuniformly magnetized medium, the total energy depends on the
relative orientation of vectors $\mathbf{m}(\mathbf{r})$ at different
points. The stability of the magnetic state with respect to the variation of
magnetization density is determined by the tensor of the second derivatives 
\begin{equation}
J_{\alpha \beta }(\mathbf{r},\mathbf{r}^{\prime })=-\frac{\delta ^{2}E}{%
\delta m_{\alpha }(\mathbf{r})\delta m_{\beta }(\mathbf{r}^{\prime })}\mid
_{m=m_{0}}  \label{C1}
\end{equation}%
where $\alpha =x,y,z$ and $m_{0}$ is the magnetization of any stationary
state. In general, unlike the parameters of Heisenberg model, the parameters 
$J_{\alpha \beta }(\mathbf{r},\mathbf{r}^{\prime })$ depend on the magnetic
configuration and are not necessarily related to the total energy
differences between different magnetic phases. The quantity $J_{\alpha \beta
}$ is a model quantity and the definition (\ref{C1}) is not unique.

To calculate $J_{\alpha \beta }(\mathbf{r},\mathbf{r}^{\prime })$ one has to
know the variation over the spin density. The real physical perturbation
which can induce such a variation is the external magnetic field. Thus, we
have to employ the ideas of linear response and calculate the dynamic
magnetic susceptibility (DMS)\cite{GOUTIER}. In this case the variation of
magnetization is written as

\begin{equation}
\delta \mathbf{m}=\widehat{\chi }\delta \mathbf{B}_{ext},  \label{C2}
\end{equation}%
where $\widehat{\chi }$ is the (non-local) DMS which can be presented as 
\begin{equation}
\widehat{\chi }=\frac{\delta \mathbf{m}}{\delta \mathbf{B}_{ext}}=-\widehat{%
\chi }\frac{\delta ^{2}E}{\delta \mathbf{m}(\mathbf{r})\delta \mathbf{m}(%
\mathbf{r}^{\prime })}\widehat{\chi }=\widehat{\chi }\widehat{J}(\mathbf{r},%
\mathbf{r}^{\prime })\widehat{\chi },  \label{C3}
\end{equation}%
Therefore $\widehat{J}=\widehat{\chi }^{-1}.$This definition of the
effective exchange is nothing but the well-known connection between the
inverse static susceptibility and the parameter $\Phi _{2}$ in the Landau
theory of phase transitions (T.Moriya also uses the inverse susceptibility
in the Heisenberg model; see Ref.\cite{MORIYA}). The goal of this paper is
to consider the relationship between this definition and those currently
used in magnetism research. Also the corresponding modification of the
adiabatic SW spectra will be analyzed.

Let us generalize this concept and introduce the frequency dependent
exchange coupling $\widehat{J}(\mathbf{r,r}^{\prime \prime },\omega )$ as 
\begin{equation}
\int d\mathbf{r}^{\prime \prime }\widehat{J}(\mathbf{r,r}^{\prime \prime
},\omega )\widehat{\chi }(\mathbf{r}^{\prime \prime }\mathbf{,r}^{\prime
},\omega )=\delta \left( \mathbf{r}-\mathbf{r}^{\prime }\right) .  \label{C4}
\end{equation}

The parameter $\widehat{J}(\mathbf{r,r}^{\prime },\omega )$ (below we will
omit matrix notations where it is not important) is the full exchange
containing all enhancement effects. Some approximate expressions for $J(%
\mathbf{q,}\omega )$ in a long wave approximation have been obtained in Ref.%
\cite{REVIEW}. Unfortunately, the exact calculation of DMS in the case of
interacting particles is not possible at the moment and any practical
evaluations of $J(\mathbf{r,r}^{\prime },\omega )$ require certain
simplifying assumptions.

Using linear response for interacting electrons for the inverse DMS one can
decompose the total exchange as follows: 
\begin{equation}
J(\mathbf{r,r}^{\prime },\omega )=J^{0}(\mathbf{r,r}^{\prime },\omega
)-I_{xc}(\mathbf{r,r}^{\prime },\omega )  \label{c99}
\end{equation}%
Here $J^{0}(\mathbf{r,r}^{\prime },\omega )$ is the 'bare' exchange coupling
(obtained, for instance, from the Kohn-Sham wave functions) whereas $I_{xc}(%
\mathbf{r,r}^{\prime },\omega )$ is the exchange and correlation tensor that
connects $\mathbf{B}_{xc}$ to the induced magnetization $\mathbf{B}_{xc}(%
\mathbf{r,}\omega )=\int d\mathbf{r}^{\prime }I_{xc}(\mathbf{r,r}^{\prime
},\omega )\mathbf{m}\left( \mathbf{r}^{\prime },\omega \right) .$

Eq.\ref{c99} represents a very convenient way to add enhancement effects to
the exchange coupling parameter. In the absence of an explicit form for $%
I_{xc}$ from many-body theory, Eq.\ref{c99} must be considered as a formal
definition of the frequency dependent Stoner parameter\ $I_{xc}(\mathbf{r,r}%
^{\prime },\omega ).$ It becomes practical only when the translational
invariance of a perfect crystal $J(\mathbf{r,r}^{\prime },\omega )=J$ $(%
\mathbf{r-r}^{\prime },\omega )$ and the condition $I_{xc}(\mathbf{r,r}%
^{\prime },\omega )=I_{xc}$ $(\mathbf{r-r}^{\prime },0)=I$ are assumed. Then
the condition of the SW existence is written as%
\begin{equation}
J^{0}(\mathbf{q,}\omega )=I_{xc}.  \label{c95}
\end{equation}%
One has to emphasize that Eq.(\ref{c99}) is suitable for full-potential band
structure treatment and requires a \textit{complete} basis set for its
computation for the real materials. Also, below we will consider only the
ferromagnetic case and will limit our consideration to analyzing the effects
related to $\chi ^{+-}$ component. For arbitrary magnetic ordering, the
matrix $\widehat{\chi }(\mathbf{r,r}^{\prime },\omega )$ is a 4x4 matrix and
its $\chi ^{+-}$ component can not be easily separated from $\chi ^{zz}$ and
charge components\cite{GOUTIER}.

To illustrate how the parameter $J$ enters the spin dynamics (SD) equation
of motion (EOM), let us consider the adiabatic limit of Eq.\ref{c99}. In
this case a following density functional torque equation is valid%
\begin{equation}
\frac{d\mathbf{m}\left( \mathbf{r},t\right) }{dt}=\gamma \mathbf{m}(\mathbf{%
r,}t)\times \mathbf{B}(\mathbf{r})  \label{c101}
\end{equation}%
where $\mathbf{B}$ the total field acting on the electronic spin at point $%
\mathbf{r}$ and $\gamma $ is gyromagnetic ratio. Not all components of $%
\mathbf{m}$ are independent in Eq.\ref{c101} because of the relation $%
\mathbf{m\cdot }d\mathbf{m}/dt=0.$

For the adiabatic SD Eq.(\ref{c101}) can be solved for several simplest
magnetic orderings by linearizing this EOM analogously to the rigid spin
approximation (RSA) case\cite{SD}. For instance, for ferromagnet (FM) we
obtain 
\begin{equation}
\widehat{\omega }_{\mathbf{q}}=m\left[ \chi _{\mathbf{q}}^{-1}K_{\mathbf{q}%
}-\chi _{0}{}^{-1}\right] .  \label{c13}
\end{equation}%
where $\chi _{\mathbf{q}}$ is the Fourier transform of the 'bare' static
susceptibility $\chi (\mathbf{r}-\mathbf{r}^{\prime },\omega =0)$ in the
case of a periodic system. $K_{\mathbf{q}}$ is a 'kinetic gradient' matrix
which takes into account the difference in spatial gradients of wave
functions for the different spin direction\cite{SDNI}. This result for the
adiabatic SW spectra is obtained here in the non-local case without using
RSA for the shape of magnetic perturbation. For more complicated magnetic
structures (spirals, and etc) the corresponding dispersion laws may be
obtained using a rotating coordinate system. The above results were obtained
assuming that the decay of SW is absent (adiabatic SD); otherwise these
equations are exact (in linear regime) and treat both short and long
wavelength scales on an equal footing. If a magnetic field of relativistic
origin is included, then in Eq.(\ref{c13}) one can add the energy of the
free precession (rotation of the magnetization of the whole crystal).

To analyze the relationship between the dispersion law (\ref{c13}) and that%
\cite{KORENMAN} commonly used in the DFT, $\omega _{\mathbf{q}}^{l}=mI\left[
\chi {}_{0}-\chi _{\mathbf{q}}\right] I$, we assume that the ratio%
\begin{equation}
\Delta =Tr\widehat{\Delta }=Tr\left( \chi _{\mathbf{q}}{}-\chi _{0}\right)
\chi _{0}^{-1}\approx \omega _{\mathbf{q}}^{l}/mI  \label{cn}
\end{equation}%
is small (long-wave approximation) and $K_{\mathbf{q}}=1$ (effective
bandwidths for the different spin directions are equal). Then, by expanding
Eq.\ref{c13} over the parameter $\Delta ,$ we obtain the desired result:%
\begin{eqnarray}
\widehat{\omega }_{\mathbf{q}} &=&m\left( \chi _{\mathbf{q}}^{-1}-\chi
_{0}{}^{-1}\right) =m\chi _{0}^{-1}\widehat{\Delta }\left( 1-\widehat{\Delta 
}\right) ^{-1}  \label{c15} \\
&=&\widehat{\omega }_{\mathbf{q}}^{l}\left( 1-\widehat{\omega }_{\mathbf{q}%
}^{l}\left( \widehat{I}m\right) ^{-1}\right) ^{-1}\simeq m\chi
_{0}^{-1}\left( \chi {}_{0}-\chi _{\mathbf{q}}\right) \chi _{0}^{-1}  \notag
\\
&=&m\left( J_{0}^{l}-J_{\mathbf{q}}^{l}{}\right)  \notag
\end{eqnarray}%
or%
\begin{equation}
J_{\mathbf{q}}=J_{0}\left[ 1+\widehat{\Delta }\left( 1-\widehat{\Delta }%
\right) ^{-1}\right] =J_{0}\left[ 1+\widehat{\Delta }+\widehat{\Delta }%
^{2}+..\right] ,  \label{c190}
\end{equation}%
where the susceptibility has a matrix structure and%
\begin{equation}
J_{\mathbf{q}}^{l}=J_{0}\left( 1+\widehat{\Delta }\right) =\chi
_{0}^{-1}\chi _{\mathbf{q}}\chi _{0}^{-1}=I\chi _{\mathbf{q}}I=-I\frac{%
\partial ^{2}E}{\partial \mathbf{B}_{tot}\partial \mathbf{B}_{tot}}I
\label{c166}
\end{equation}%
is the matrix of the exchange parameter in the local (long-wave)
approximation. Eq.(\ref{c166}) is the widely accepted definition of the
exchange coupling parameter which was introduced in Ref.\cite%
{MORIYA,LIU,KORENMAN} \ and used in different modifications in Ref.\cite%
{OGU,LIXT,JMMM,REVIEW}. Eq.\ref{c13} and the relations (\ref{c15},\ref{c190}%
) are the main results of this paper. Below we will apply this result to the
calculations of the exchange parameters and adiabatic SW spectra in Fe, Ni,
and Gd using the multiple scattering technique.

From the computational point of view two alternatives can be used. First one
is a calculation of the total energy of the system with an external field
included and the second one is to use of so-called 'local force' theorem
which does not require self-consistency and the total energy calculations.

In Ref.\cite{LIXT} a general 'local force' theorem was formulated for
magnetic perturbations and a convenient multiple scattering technique was
suggested for calculations of $\chi _{xc}=-\partial ^{2}E/\partial \mathbf{B}%
_{xc}\partial \mathbf{B}_{xc}\mathbf{,}$ where a role of $\mathbf{B}_{xc}$
belongs to one-site scattering matrix $t^{-1}$. We stress that this theorem
can be used for the calculations of both the magnetic susceptibility $\chi
^{+-}$ and the effective exchange $J=-\partial ^{2}E/\partial \mathbf{m}%
\partial \mathbf{m}$ \textit{without} any modifications of the original
theorem. This statement can be also seen as a trivial result of the well
known linear response theory.

The usual assumption of weak enhancement (see Eq.\ref{c99}) has always been
used previously. This is a reasonable approximation for the SW spectrum,
because if we assume usual $I_{xc}(\mathbf{q})=I,$ then the important
equality $J(\mathbf{q})-J(0)=J^{0}(\mathbf{q})-J^{0}(0)$ is valid. As a
consequence, the spectrum of elementary excitations is not affected by
exchange-correlation enhancement effects. So, the definition (\ref{c166}) is
directly related to several very strong approximations: RSA, smallness of SW
dispersion compared to the effective exchange splitting (atomic limit), and
small dispersion of $I_{xc}(\mathbf{q,}\omega )$.

From the formal point of view, the RSA was directly contained in the way
that the local force theorem was previously used. While the exact
formulation of this theorem contains the first functional \textit{variation}
of the total energy, in RSA one has to use the corresponding first \textit{%
derivative} 
\begin{equation}
\frac{\delta E}{\delta \mathbf{m}\left( \mathbf{R}_{i}+\mathbf{r}\right) }%
\approx \frac{\partial E}{\partial \mathbf{m}_{i}}  \label{c167}
\end{equation}%
with all gradient terms (large $\mathbf{q}$ vectors) omitted. Here $i$ is
index of the atomic site.

The second approximation $\left( \chi _{\mathbf{q}}{}-\chi _{0}\right) \chi
_{0}^{-1}<<1$ also removes certain degrees of freedom with large $\mathbf{q}$
which are present in the Brillouin zone. Analogous one site approximations
have been employed in modern dynamic mean field techniques\cite{DMF},
self-consistent GW scheme\cite{US} and the coherent potential approximation%
\cite{OGU}. Such a restriction affects the spin-spin correlation function
and in turn the dynamic and thermal properties of magnets. It is true that
(as one can easily obtain from Eq.\ref{c190}), such a long wave
approximation can safely be used for SW stiffness calculations.

The term $I_{xc}\left( \mathbf{q}\right) $ is usually short-ranged, which
coincides with the corresponding smallness criteria in real space $\chi
_{ij}{}/\chi _{ii}<<1$ (where $\chi _{ij}$ is a Fourier transform of $\chi _{%
\mathbf{q}}$ and $\chi _{ii}$ is the on-site susceptibility), hence the
local approximation $I_{xc}(\mathbf{q})=I$ may affect the exchange between
the nearest atoms only.

Let us quantitatively discuss the range of applicability of the above
formalism for real magnetic systems. To perform realistic calculations we
use multiple scattering theory (see, for instance\cite{REVIEW} and
references therein) to obtain 'bare' $J_{\mathbf{q}}.$ In this theory a key
equation is

\begin{equation}
\tau (\varepsilon )=[P(\varepsilon )-S]^{-1}  \label{E3}
\end{equation}%
where $\tau (\varepsilon )$ is scattering path operator, $P(\varepsilon )$
is the inverse of the one-site scattering matrix and $S$ is the matrix of
structure constants. It is convenient to separate spin structure explicitly:

\begin{equation}
\tau (\varepsilon )=T_{0}(\varepsilon )+\mathbf{T}(\varepsilon )\mathbf{%
\sigma ,}P(\varepsilon )=p(\varepsilon )+\mathbf{p}(\varepsilon )\mathbf{%
\sigma .}  \label{ee5}
\end{equation}%
Then the total energy variation with respect to the deviation of magnetic
field at the site $i$ can be presented in the static linear response scheme
as\cite{REVIEW}%
\begin{equation}
\delta E=-\frac{2}{\pi }\int^{\varepsilon _{F}}d\varepsilon \mathrm{ImTr}%
_{L}\left\{ \delta \mathbf{p}_{0}\mathbf{T}_{00}\right\} ,  \label{c34}
\end{equation}%
where $\mathbf{T}_{00}$ is a matrix element of vector component of the full
scattering matrix at site $0.$ Using the sum rule for $\tau $ matrix
introduced in Ref.\cite{REVIEW}, we can obtain the following relation for
the 'bare' exchange coupling 
\begin{eqnarray}
J_{\mathbf{q}}^{+-} &=&\frac{1}{4\pi }\int^{\varepsilon _{F}}d\varepsilon 
\mathrm{ImTr}_{L}\left( T^{\uparrow }-T^{\downarrow }\right) _{00}\left[
\int d\mathbf{k}T_{\mathbf{k}}^{\uparrow }T_{\mathbf{k}+\mathbf{q}%
}^{\downarrow }\right] ^{-1}\left( T^{\uparrow }-T^{\downarrow }\right)
_{00}=  \label{c168} \\
&&\frac{1}{4\pi }\int^{\varepsilon _{F}}d\varepsilon \mathrm{ImTr}_{L}\left(
T^{\uparrow }-T^{\downarrow }\right) _{00}\left[ \chi \right] _{\mathbf{q}%
}^{-1}\left( T^{\uparrow }-T^{\downarrow }\right) _{00},  \notag
\end{eqnarray}%
This expression properly takes into account the different energy dispersion
for different spins in band magnets. Eq.\ref{c168} can be directly
generalized for the non-collinear or spin spiral ordering. In this case
general 4x4 supermatrix $\chi $ should be build and inverted. Eq.\ref{c168}
can be seen as a generalization of long-wave approximation result\cite{LIXT}
for the magnets with arbitrary degree of spin localization. Below we used
the local density approximation with a linear muffin-tin orbital technique
(LMTO) in the atomic sphere approximation\cite{Mark} to calculate the
effective exchange according to Eq.\ref{c168}, the SW spectrum 
\begin{equation}
\omega _{\mathbf{q}}=\left[ J_{\mathbf{q}}-J_{0}{}\right] /m,  \label{c169}
\end{equation}%
and the SW spectrum of the localized model as 
\begin{equation}
\omega _{\mathbf{q}}^{l}=\left[ J_{0}^{l}-J_{\mathbf{q}}^{l}{}\right] /m,
\label{c170}
\end{equation}%
with long-wave $J_{\mathbf{q}}^{l}$ determined in Ref.\cite{LIXT,REVIEW}. Eq.%
\ref{c168} is a static analog of Eq.\ref{C4}, implemented in multiple
scattering technique. Both the tetrahedron scheme and fast Fourier transform
methods have been used for the Brillouin zone integration\cite{REVIEW}. The
matrix form of Eq.\ref{c168} and a proper symmetrization of the exchange
matrix from Eq.(\ref{c190}) are essential ingredients of such calculations.
The computation of the inverse of $\widehat{\omega }_{\mathbf{q}}^{0}$ with
only $s,p$ and $d$ \ basis set is numerically more stable compared to $\chi
_{\mathbf{q}}^{-1}$, but it is still affected by the type of approximation
used for the effective tight-binding LMTO Hamiltonian (see below). Three FM
systems with entirely different degree of localization of the local moment
were considered: Gd, Ni and Fe. An important observation is that in Gd
(highly localized moments, small SW dispersion) the approximation based on
the assumption $\Delta =Tr\left( \chi _{\mathbf{q}}{}-\chi _{0}\right) \chi
_{0}^{-1}<<1$ is completely fulfilled ($\Delta $ at the zone$\ $boundary is
less then $0.01$), whereas in Ni (moderately localized moments, large spin
wave dispersion) it is not valid at all ($\Delta \left( Y\right) \simeq 0.6$)%
$,$ so that the corresponding matrix estimations using Eq.(\ref{c15})
demonstrate a strong enhancement of SW spectra at larger $\mathbf{q}$ (about
50\% at $Y$ point with two center approximation of LMTO\cite{Mark} and 70\%
using third order Hamiltonian with combined correction terms). Fe is an
intermediate case where the maximum of $\Delta $ is 0.30. This result
indicates that the previous spin spiral calculations, where the derivative
with respect to exchange-correlation field instead of local moment was used,
the dispersion at finite $\mathbf{q}$ was underestimated. However, in the
small $\mathbf{q}$ regime (SW stiffness), as we have shown above (see Eq.\ref%
{c15}), the old results are perfectly correct.

It is important to analyze the effective exchange coupling between atoms in
real space. Our results indicated that in Fe and Ni a main contribution
coming from the renormalization of the first nearest neighbour exchange, so
that in bcc Fe it is increased from $J_{01}^{l}$=16.6 meV to $J_{01}$=19.4
meV, whereas in fcc Ni $J_{01}^{l}$=2.7 meV to $J_{01}$=8.3 meV.

Such a clear difference in the results indicate that the removal of the
long-wave approximation can serve as an indicator of the degree of
localization (parameter $\left( \chi _{\mathbf{q}}{}-\chi _{0}\right) \chi
_{0}^{-1}$above) in different metallic magnets. In Fe and Ni, for instance,
it at least partially explains why long wave mean field (MF) estimates
predict such a small $T_{c}$ in FM Ni: 300-350K in Ref.\cite%
{KORENMAN,REVIEW,Mark}, 340 K in present calculations, while experimental
result is 630 K. First of all, long-wave approximation is suitable for such
a 'localized' system as Fe, and the corresponding change in the nearest
neigbor $J_{01}$ is relatively small (correspondingly the increase of $T_{c}$
is expected to be small). FM Ni represents a rather itinerant system and any
local approach (long wave approximation in particular) might produce a large
error. In our case, a large increase in $J_{01}$ for Ni indicates that the
traditional MF approach (or any other approach which is based on 'no
short-range order' assumption) is not applicable for the itinerant systems
in general, predicting very high $T_{c}$ (above 1000K). Such a number is
also not consistent with the local density approach because latter does not
allow to have $T_{c}$ in fcc Ni larger then 500-540K (600-640K using
gradient corrections) and this number can be considered only as an indicator
of non-applicability of MF approach.

So, the formalism above (see Eq.(\ref{c168})) tends to raise $T_{c}$ by
inhibiting short-wavelength fluctuations and thus increasing the short-range
order. Since this estimate already includes classically all wavelengths
(except longitudinal) in the MF approximation and the final $T_{c}$ in Ni is
expected to be very high, we have thus far first indication that other
important mechanisms (with much 'softer' modes) are still missing in the
modern finite-temperature magnetism theories. Our preliminary SD simulations%
\cite{SDNI} revealed the importance of the short range order effects in Ni
and we believe that the proper definition of pairwise and non-pairwise
interactions, which include both short and long range wave lengths
correctly, is vitally important for the reliable description of the
itinerant magnetism at finite temperatures in 3d metals.

In summary, we presented a technique for the calculation of the effective
exchange parameters without any assumptions about the form of magnetization
density and the degree of magnetic order. Our results for elementary magnets
(Fe, Ni and Gd) indicated that the most significant improvement is obtained
for exchange coupling between nearest magnetic atoms and for SW spectrum at
finite wave vectors. The large error of previous long-wave approximation is
determined for fcc Ni. This result indicates also that some important
effects (which are essential for the correct description of $T_{c}$ in
itinerant magnets) are still \textit{missing} from the current magnetism
theories when the nearest neighbors $J_{ij}$ parameters determined at T=0 K
are being used at the finite temperatures. In Fe and especially in Gd the
proposed technique gives the result similar to localized model result, as
expected.

Overall, we expect that the emphasized above approach for the exchange
coupling and SW spectra Eq.(\ref{c13},\ref{c190},\ref{c168}) will
significantly improve the description of the nearest neighbors interaction
or large $\mathbf{q}$ part of SW spectra in any magnet with not \textit{fully%
} localized magnetism ($m<2\mu _{B\text{ }}$and metallic $r_{s}$). We expect
an increasing importance of this formalism at the finite temperatures and in
systems with strong magnetic short range order.

The author thank N. E. Zein and K. D. Belashchenko for many stimulating
discussions and P. Bruno for the opportunity to present the results of this
paper at the seminar in his group in Halle, in May 2002. This manuscript has
been authored by Iowa State University of Science and Technology under
Contract No. W-7405-ENG-82 with the U.S. Department of Energy.

\end{document}